\documentclass[fleqn,12pt,twoside]{article}
\usepackage[headings]{espcrc1}
\usepackage{wrapfig}

\readRCS
$Id: espcrc1.tex,v 1.2 2004/02/24 11:22:11 spepping Exp $
\usepackage{graphicx}
\usepackage[figuresright]{rotating}

\newcommand{\AmS}{{\protect\the\textfont2
  A\kern-.1667em\lower.5ex\hbox{M}\kern-.125emS}}

\title{A Hadron Blind Detector for the PHENIX Experiment at RHIC}

\author{I.~Ravinovich\address[WIS]{\vspace{-0.3cm} Weizmann Institue of Science, Rehovot 76100, Israel},
B.~Azmoun\address[BNL]{\vspace{-0.3cm} Brookhaven National Laboratory, Upton NY 11973-5000, USA},
L.~Baksay\address[FIT]{\vspace{-0.3cm} Florida Institue of Technology, Melbourne FL 32901, USA},
C.~Y.~Chi\address[CU]{\vspace{-0.3cm} Columbia University, New York, NY 10027, USA},
A. Drees\address[SB]{\vspace{-0.3cm} Stony Brook University, Stony Brook, SUNY, NY 11794-3400, USA},
A.~Dubey\addressmark[WIS], \\
Z.~Fraenkel\addressmark[WIS],
J.~Franz\addressmark[SB],
M.~Grosse-Perdekamp\address[UI]{\vspace{-0.3cm} University of Illinois, Urbana-Champaign, Urbana, IL61801, USA},
H.~Hamagaki\address[UT]{\vspace{-0.3cm} University of Tokyo, Tokyo 162-0044, Japan},
J.~Harder\addressmark[BNL], \\
T.~Hemmick\addressmark[SB],
M.~Hohlmann\addressmark[FIT],
R.~Hutter\addressmark[SB],
B.~Jacak\addressmark[SB],
D.~Kawall\address[UM]{\vspace{-0.3cm} University of Massachusets, Amherst, MA 01003, USA},
A.~Kozlov\addressmark[WIS], \\
D.~Lynch\addressmark[BNL],
M.~McCumber\addressmark[SB],
A.~Milov\addressmark[BNL],
M.~Naglis\addressmark[WIS],
P.~O'Connor\addressmark[BNL],
S.~Oda\addressmark[UT], \\
K.~Ozawa\addressmark[UT],
R.~Pisani\addressmark[BNL],
V.~Radeka\addressmark[BNL],
S.~Rembeczki\addressmark[FIT],
D.~Sharma\addressmark[WIS],
A.~Sickles\addressmark[SB], \\
A.~Toia\addressmark[SB],
I.~Tserruya\addressmark[WIS],
C.~Woody\addressmark[BNL],
B.~Yu\addressmark[BNL]
}

\runtitle{A Hadron Blind Detector for the PHENIX Experiment at RHIC}
\runauthor{I. Ravinovich for the PHENIX Collaboration}

\begin{document}

\maketitle

\begin{abstract}
A novel Hadron Blind Detector (HBD) has been developed for an upgrade of the PHENIX
experiment at RHIC. The HBD will allow a precise measurement of electron-positron pairs
from the decay of the light vector mesons and the low-mass pair continuum in heavy-ion collisions.
The detector consists of a 50 cm long radiator filled with pure CF$_4$ and directly coupled
in a windowless configuration to a triple Gas Electron Multiplier (GEM) detector with a CsI
photocathode evaporated on the top face of the first GEM foil.
\end{abstract}

\section{Introduction}

In this paper we present a summary of the R$\&$D results and expected performance of a Hadron Blind
Detector (HBD) which has been developed as an upgrade of the PHENIX detector at the
Relativistic Heavy Ion Collider (RHIC) at BNL. The HBD will allow the measurement of
electron-positron pairs from the decay of the light vector mesons ($\rho$, $\omega$ and $\phi$)
and the low-mass ($m_{e^+e^-} \leq$ 1 GeV/c$^2$) pair continuum in Au+Au collisions.
Dileptons are valuable probes to diagnose the hot and dense matter formed in relativistic
heavy-ion collisions. The physics potential of this probe is fully confirmed by results from lower energy
experiments. The most prominent result is the strong enhancement of low-mass electron pairs observed
by the CERES experiment \cite{ceres} at the CERN SPS and recently confirmed by the NA60
results \cite{NA60}. This enhancement triggered a wealth of theoretical activity and was
quantitatively reproduced only by invoking the thermal radiation from a high density hadron gas
($\pi^+\pi^- \rightarrow e^+e^-$)
with in-medium modification of the intermediate $\rho$ meson which could be linked to chiral symmetry
restoration \cite{rapp-brown}.

The measurement of electron pairs is however a very challenging one. The main difficulty arises from the 
huge combinatorial background, i.e. uncorrelated pairs formed by tracks from unrecognized
$\gamma$ conversions and $\pi^0$ Dalitz decays. For example, the analysis of Run-4 data taken with the
present PHENIX configuration \cite{toia} shows that the signal to background ratio, S/B, at the invariant
mass of m $\sim$~500~MeV/c$^2$ is approximately 1/500, making the measurement of the low-mass pair
continuum practically impossible. An upgrade of the PHENIX detector is therefore necessary for this
measurement. The present paper is focussed on the comprehensive R\&D program that we have carried
out over the past two years to develop such an upgrade \cite{NIM1,NIM2}.

\section{The HBD concept}

\begin{figure}[htb]
\vspace{-1cm}
\begin{minipage}[t]{80mm}
  \includegraphics[keepaspectratio=true, width = 7cm]{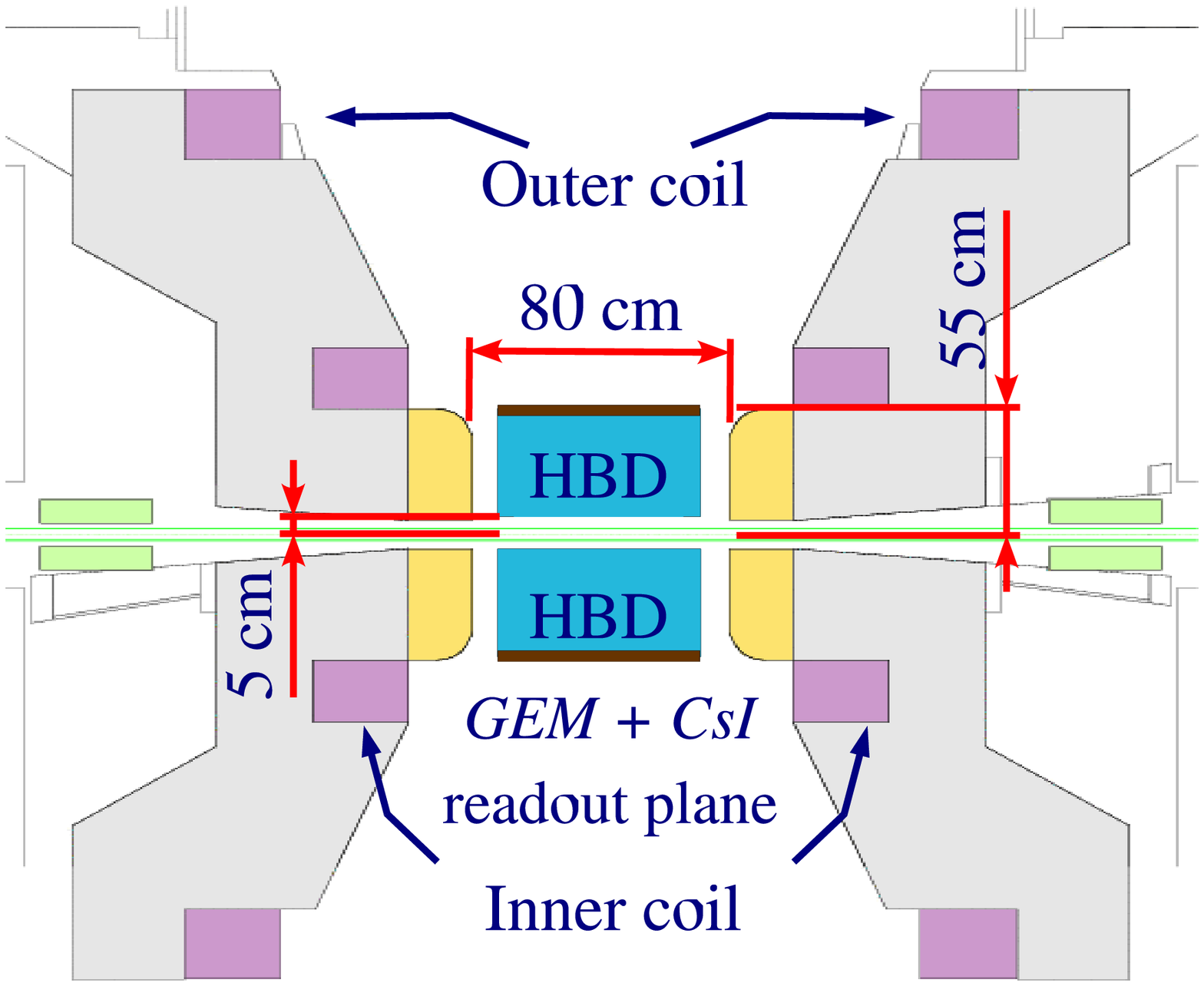}
\vspace{-1.25cm}
  \caption{Layout of the inner part of the PHENIX detector showing the location of the
          HBD and the inner coil.}
\label{fig:hbdinphnx}
\end{minipage}
\hspace{\fill}
\begin{minipage}[t]{75mm}
\vspace{-6.55cm}
  \includegraphics[keepaspectratio=true, width = 5.5cm]{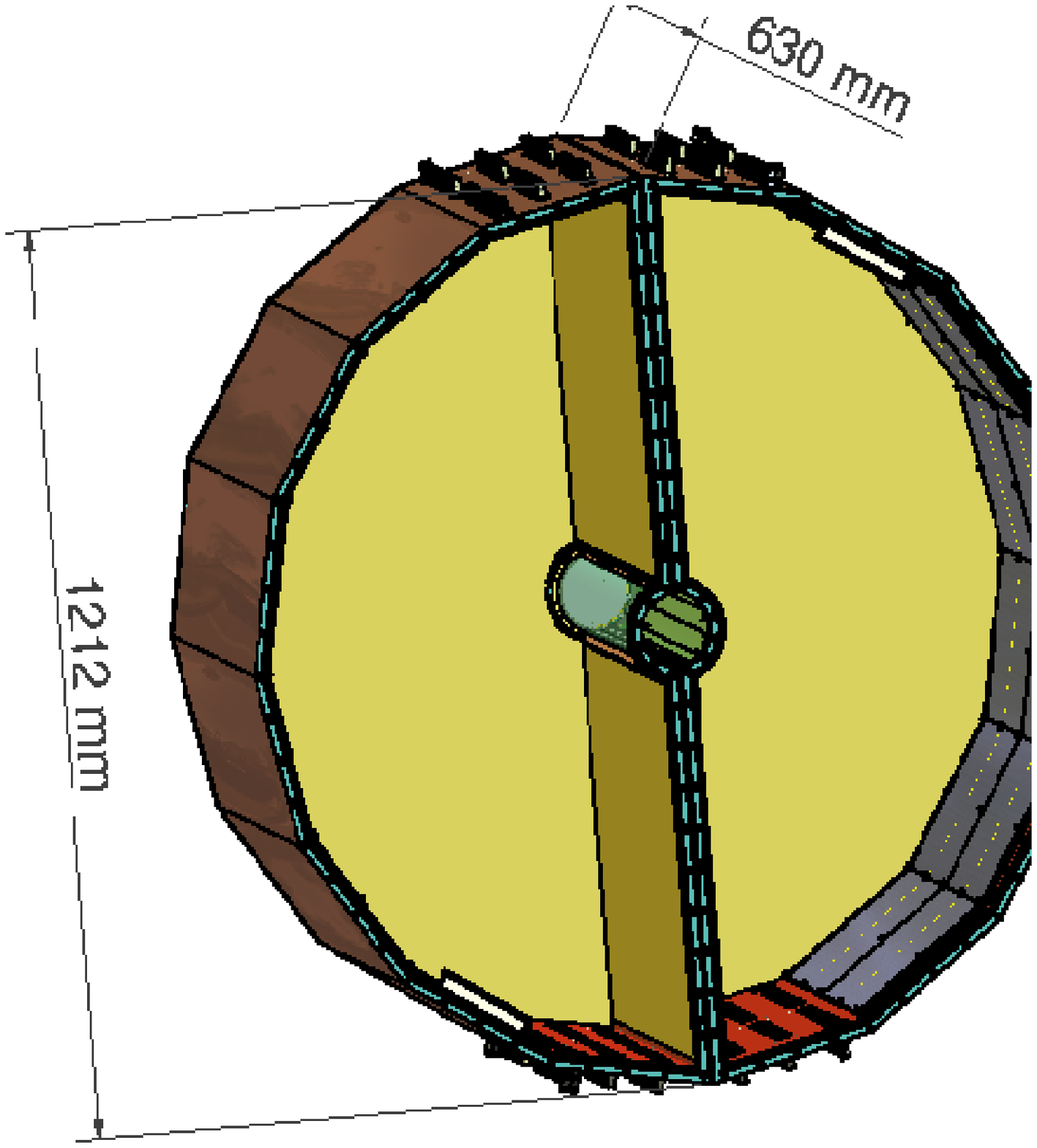}
\vspace{-0.65cm}
  \caption{3-d view of the HBD final design.}
\label{fig:hbd-3d_qm2005}
\end{minipage}
\vspace{-1cm}
\end{figure}

The upgrade consists of two elements: a) installation (foreseen in the original design of PHENIX) of an
inner coil which results in an almost field-free region close to the vertex, extending out to $\sim$ 60 cm
in the radial direction; b) the major and challenging element of the upgrade is a hadron-blind detector
(HBD) located in this field free region. Fig.~\ref{fig:hbdinphnx} shows the layout of the inner part of the
PHENIX detector together with the location of the coils and the proposed HBD. The main task of the
HBD is to recognize and reject $\gamma$ conversions and $\pi^o$ Dalitz decays. The strategy is to
exploit the fact that the opening angle of electron pairs from these sources is very small compared to
the pairs from light vector mesons. In the field-free region, this angle is preserved  and by applying an
opening angle cut one can reject more than 90\% of the conversions and $\pi^o$ Dalitz decays, while
preserving most of the signal.

Based on Monte Carlo simulations, the main HBD specifications are: electron identification with a very
high efficiency ($>$ 90$\%$), double hit recognition at a comparable level and moderate $\pi$ rejection
factor of $\sim$100. We analyzed possible realizations of the HBD detector and finally adopted the
following scheme: a windowless \v{C}erenkov detector, operated with pure CF$_4$ in a proximity
focus configuration and directly coupled to a triple Gas Electron Multiplier (GEM) \cite{Sauli_GEM}
detector with a CsI photocathode evaporated on the top face of the first GEM foil and with pad readout.

\section{Summary of R$\&$D results}

The concept described in the previous section exhibits a number of very attractive features:
a) the choice of CF$_4$ both as radiator and detector gas in a windowless geometry results in a very
large bandwidth (6-11.5 eV) and consequently in a very large figure of merit
N$_0$ ($\sim$800~cm$^{-1}$). With these unprecedented numbers, the
number of photoelectrons N$_{pe}$ is expected to be of the order of 35 for a 50 cm long radiator;
b) the readout scheme foresees the detection of the \v{C}erenkov photoelectrons in a pad plane with
hexagonal pads of size slightly smaller than the blob size ($\sim$10 cm$^2$) such that the probability
of a single-pad hit by an electron entering the HBD is negligibly small. On the other hand, whenever a
hadron gives a signal in the HBD, it will produce a single pad hit with an almost 100\% probability. This
will provide a strong handle in the hadron rejection factor of the HBD;
c) the relatively large pad size results also in a low granularity and therefore a low cost detector.
In addition, since the photoelectrons produced by a single electron will be distributed between at most
three pads, one can expect a primary charge of at least 10 electrons/pad, allowing operation of the
detector at a relatively moderate gain of a few times 10$^3$. This is a crucial advantage for the stable
operation of a UV photon detector. However, many elements of the proposed HBD were new and had
not been tested before in the laboratory. A number of issues and questions were raised which required
dedicated investigation. To address these questions and to demonstrate the concept  validity, a
comprehensive R\&D program was needed. The results of this effort are summarized below
\cite{NIM1,NIM2}:
\vspace{-0.25cm}
\begin{itemize}
\itemsep=-3pt
\item We have shown that a triple GEM detector with a reflective CsI photocathode operates in a stable
mode at gains up to 10$^4$ in pure CF$_4$.

\item A charge saturation effect occurring in CF$_4$ when the total charge in the avalanche reaches
4$\times$10$^7$~e makes the HBD relatively robust against discharges.

\item We carried out a test of a triple GEM detector operated with pure CF$_4$ at the proposed location
inside the PHENIX central arm spectrometer. The detector performed smoothly in the presence of Au+Au
collisions exhibiting no discharges or gain instabilities.

\item We have studied the basic parameters which determine the HBD performance. In particular, we
measured the device response to mip's and to electrons. Large hadron rejection factors, well in excess
of 100, can be achieved while preserving an electron detection efficiency larger than 90\%.

\item We confirmed measurements of the CsI quantum efficiency over the bandwidth 6-8.3~eV and 
extended them up to 10.3~eV. Extrapolation to the expected operational bandwidth of the device
(6-11.5 eV) gives a figure of merit N$_0$=822 cm$^{-1}$.

\item Aging studies of the GEM foils as well as the CsI photocathode revealed that there is no significant
deterioration of the detector for irradiation levels corresponding to $\sim$~10 years of normal PHENIX
operation at RHIC.

\end{itemize}
\vspace{-0.25cm}
These results demonstrated the validity of the proposed HBD concept and paved the way to the
incorporation of such a device in the PHENIX experiment.


\begin{wrapfigure}{r}{60mm}
  \vspace{-20mm}
  \centering
  \includegraphics[keepaspectratio=true, width=55mm]{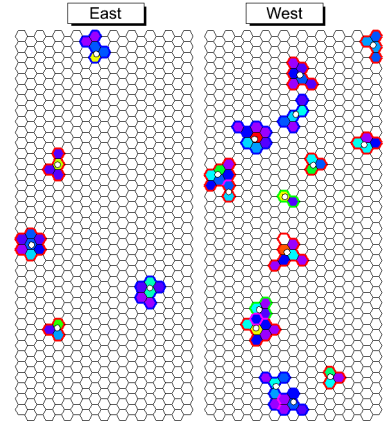}
  \vspace{0mm}
  \caption{HBD event display of a central HIJING event.}
  \label{fig:event_display}
  \vspace{-5mm}
\end{wrapfigure}

\vspace{5mm}
\section{Detector design and system performance}

The mechanical design of the HBD has been developed and construction procedures have been exercised
and optimized on a full-scale prototype. The HBD consists of two identical arms
(see Fig.~\ref{fig:hbd-3d_qm2005}), each one covering 135$^o$ in azimuth and $\pm$0.45 units of pseudorapidity.
In each arm the detector element is subdivided in 12 detector modules, 6 along the $\phi$ axis $\times$ 2
along the z axis, with a module size of $\sim 23 \times 27$ cm$^2$. The signals are collected in the anode
plane consisting of 1152 hexagonal pads in each arm. The detector vessel has a polygonal shape made
of panels glued together.

The full-scale prototype design has been integrated into the PHENIX standard simulation package. 
Realistic simulations have been performed using HIJING central Au+Au collisions at $\sqrt{s_{NN}}=200$
GeV with embedded $\phi$ mesons. We chose a very narrow centrality window: impact parameter
b~$<$ 2~fm (top 2$\%$) which corresponds to an average charged particle rapidity density of
dN$_{ch}$/dy~$=$~940. Fig.~\ref{fig:event_display} shows the reconstructed clusters after clean-up
of such a central collision. The clean-up is optimized for electron identification and hadron rejection.
It requires:
a) a pad amplitude $>1$~p.e.;
b) a cluster amplitude$ >$~20~p.e.;
c) a cluster size with  N$_{pads}\geq$~2.
Out of the huge number of hadrons ($\sim$400) going through the HBD only two are reconstructed
demostrating excellent hadron rejection. Electrons reconstructed in the central arms and matched
to a cluster in the HBD are rejected as
a) likely conversions if the cluster amplitude is larger than 60 p.e.;
b) likely $\pi^o$ Dalitz decays if there is another electron cluster in the HBD within a distance of 200~mrad.
For the mass region around the $\phi$ meson, this results in a reduction of the combinatorial background
from these two sources by a factor larger than 100, compared to the present performance without the
HBD. At this level of rejection the quality of the low-mass pair measurement is not anymore limited by
this background but rather by the combinatorial background from the semi-leptonic decays of charmed
mesons.

The HBD is presently under construction with installation and commisioning foreseen in 2006.



\end{document}